\documentclass[12pt,a4paper]{article}
\usepackage{authblk}
\usepackage[english]{babel}
\usepackage{verbatim}
\usepackage{amsfonts}
\usepackage{amsmath}
\usepackage{upgreek}
\usepackage[T1]{fontenc}
\usepackage{color}
\usepackage{epsfig}
\usepackage{natbib}
\usepackage{url}
\usepackage[bookmarksnumbered=true, colorlinks=true, citecolor=blue]{hyperref}
\date{}

\title{On the Conceptual Issues Surrounding the Notion of Relational Bohmian Dynamics}
\author[$\dagger$]{Antonio Vassallo}
\author[$\ddagger$]{Pui Him Ip}
\affil[$\dagger$]{University of Lausanne, Department of Philosophy, CH-1015 Lausanne}
\affil[$\ddagger$]{Magdalene College, Cambridge, CB3 0AG, U.K.}

\begin{document}

\maketitle
\begin{center}
Accepted for publication in \emph{Foundations of Physics}.
\end{center}

\pdfbookmark[1]{Abstract}{abstract}
\begin{abstract}
The paper presents a program to construct a non-relativistic relational Bohmian theory, that is, a theory of $N$ moving point-like particles that dispenses with space and time as fundamental background structures. The relational program proposed is based on the best-matching framework originally developed by Julian Barbour. In particular, the paper focuses on the conceptual problems that arise when trying to implement such a program. It is argued that pursuing a relational strategy in the Bohmian context leads to a more parsimonious ontology than that of standard Bohmian mechanics without betraying the original motivations for adopting a primitive ontology approach to quantum physics. It is also shown how a relational Bohmian approach might clarify the issue of the timelessness of the dynamics resulting from the quantization of a classical relational system of particles.\\
\\
\textbf{Keywords}: Bohmian mechanics; primitive ontology; relationalism; background independence; best-matching; shape space; time capsule.
\end{abstract}

\section{Bohmian Mechanics and Primitive Ontology}
\newcounter{contatore}
\setcounter{contatore}{0}
\newtheorem{fed}[contatore]{Conceptual Issue}

Bohmian mechanics (BM) in its modern formulation as laid down in \cite{323} is the simplest non-trivial Galilean-invariant theory of moving point-like particles. The dynamics of the theory is encoded in the following two equations:\footnote{We assume for simplicity's sake that $\hbar=1$.}
\begin{subequations}\label{BM}
\begin{equation}\label{sch}
i\frac{\partial\Psi(\mathbf{Q},t)}{\partial t}=\Bigg (-\sum_{i=1}^{N}\frac{\nabla_{i}^{2}}{2m_{i}}+V\Bigg )\Psi(\mathbf{Q},t);
\end{equation}
\begin{equation}\label{gui}
\frac{d\mathbf{Q}}{dt}=\mathbf{m}^{-1}Im\frac{\boldsymbol{\nabla}\Psi}{\Psi}(\mathbf{Q},t).
\end{equation}
\end{subequations}
Equation (\ref{sch}) is the usual time-dependent Schr\"odinger equation for a $N$-particle system interacting through a potential $V$, where the wave function $\Psi$ is defined over $\mathbb{R}^{3N}$, which is considered the configuration space of the theory: $\mathbf{Q}=\left ( \begin{array}{l}
\mathbf{q}_{1}\\
\cdots\\
\mathbf{q}_{N}
\end{array} \right )$ represents in fact a point in this space.\\
Equation (\ref{gui}) is the so-called \emph{guiding equation} for the particles, where $\boldsymbol{\nabla}=\left ( \begin{array}{l}
\boldsymbol{\nabla}_{1}\\
\cdots\\
\boldsymbol{\nabla}_{N}
\end{array} \right )$ is the ``gradient vector'', and $\mathbf{m}$ is the $N\times N$ diagonal ``mass matrix'' $\{\delta_{ij}m_{i}\}$, $m_{i}$ being the mass of the $i$-th particle.\footnote{It is important to highlight that equations (\ref{BM}) can be generalized to whatever configuration space endowed with a non-trivial Riemannian structure $g_{ij}\neq\delta_{ij}$. See \citep[][section 2]{257} for the technical details.}\\
The physical interpretation of BM is straightforward: the theory talks about $N$ massive spinless\footnote{However, the theory can be easily generalized in order to account for phenomena involving spin, as shown, for example, in \cite{380}.} point-like particles with definite positions in Euclidean $3$-space at all times; the wave function in this picture has the role of generating the vector field on the right-hand side of (\ref{gui}), and this is why it is said to ``guide'' the motion of the particles. The ``quantumness'' of BM resides in the fact that, according to (\ref{gui}), the motion of each particle is instantaneously dependent on the position of \emph{all} the other $N-1$ particles. This is the way BM implements the empirically proven non-locality of the quantum realm, that is, by virtue of (\ref{gui}) being a non-local law. Furthermore, BM is fundamentally a universal theory since (\ref{BM}) describes the dynamics of all there is in the universe - i.e. particles. However, the theory provides a consistent procedure for defining sub-configurations of particles approximately behaving as isolated quantum systems guided by an ``effective'' sub-wave function. It is exactly thanks to this fact that BM accounts for ordinary quantum measurements, thus resulting empirically equivalent to standard quantum mechanics.\footnote{See \cite{222} for a rigorous justification of these claims.}\\
As it stands, BM is the epitome of a \emph{primitive ontology approach} to quantum physics which, in a nutshell, includes all those theoretical frameworks involving a dual structure $(\mathcal{X}, \Psi)$, with $\mathcal{X}$ the primitive ontology properly said (in the present case, point-particles) and $\Psi$ being the element of the theory that dictates the dynamical evolution of $\mathcal{X}$. According to this sketch (developed, for example, in \citealp{203}), also dynamical collapse theories such as the one proposed by \cite{209} can be considered as primitive ontology approaches.\\
Usually, quantum theories involving a primitive ontology are considered a ``reaction'' to one of the most compelling conceptual problems faced by standard quantum mechanics, namely, the measurement problem. Roughly, this is because postulating the existence of fundamental ``stuff'' evolving in space and time is taken as a natural step to explain the fact that the description of quantum measurements is always given in ``classical'' terms, e.g. definite pointer positions.\\
Here we do not want to enter in any detail into the debate whether BM - or primitive ontology approaches in general - fares conceptually better than other approaches that do not postulate a primitive ontology of stuff in spacetime in accounting for quantum phenomena and their measurement. We will hence leave to the reader to go further into this debate, starting from the claim that BM solves the measurement problem of standard quantum mechanics (see, e.g., \citealp{233}). Nonetheless, three remarks are in order.\\
First of all, unlike standard quantum mechanics, BM shifts the accent from observables to beables, i.e. to stuff that is ``out there'' irrespective of the fact that a measurement operation is performed on it or not. More precisely, BM as described by (\ref{BM}) deals with point-particles which, at each time, have a definite position in physical space. This is in fact one of the essential motivations behind a primitive ontology approach: dispensing with the need for an observer. Secondly, equations (\ref{BM}) are totally consistent with a physical interpretation that does not reify in any way the wave function (by claiming, for example, that it is some sort of physical field), but just accords a law-like status to it (analogous to the status of the Hamiltonian function in phase space dynamics).\footnote{Roughly, the analogy resides in the fact that both the wave function in BM and the Hamiltonian in the phase space formulation of mechanics ``generate'' through the equations of motion a vector field in the appropriate space whose integral curves are in fact the dynamical trajectories of the physical system under scrutiny.} It is then an interesting question what metaphysical stance with respect to laws (e.g. Humeanism, modal realism) fares better in explaining the role of the wave function in BM (see, e.g., \citealp{230}). Thirdly, from what has been already said it follows that BM tells an ontological story that is clearer than that supplied by standard quantum mechanics, in that it speaks of physical systems with well-defined properties (e.g. positions) at each instant of time: in such a theory the role of superpositions, uncertainties, and collapses is merely mathematical and devoid of metaphysical import.\footnote{It is worth noting that many authors are not sympathetic to BM exactly because it brings - paraphrasing \cite{415} - classical terms into the equations. For these authors such a move amounts to forcing a ``folk'' metaphysical reading on quantum phenomena, which instead should be understood as a radical departure from our everyday picture of reality. However, this is just a declaration of metaphysical tastes since it does not entail in any way that the Bohmian approach to quantum physics, because of its intuitive character, should be empirically inadequate.}\\
Let us now focus on the very notion of primitive ontology. From the above presentation, we notice that such a designation conflates (at least) two different albeit closely related notions. Firstly, we have  ``the stuff that is guided'': in BM, just point-particles. Secondly, we have ``what there is''. It is then reasonable to constrain any proper theory falling in the scope of this approach to make these two meanings compatible.\\
According to the original aim of this kind of approaches, it seems prima facie that  ``primitive ontology'' refers to the ``stuff things are made of''  or, also, to a ``decoration of spacetime'' \citep[][p. 11]{203}. The reason for this choice is simple: if everything is made of ``primitive stuff'' with a minimal set of properties (e.g. definite positions), then a measurement can be accounted for just as some sort of ``interaction'' of elements of such primitive ontology (e.g. a pointer that points is nothing but a bunch of particles acquiring a certain spatial configuration as a result of the dynamical evolution of the global configuration of particles, including the group of particles that form the measured system). However, if this is the case, then we would immediately see that the term ``primitive ontology'' would not capture the deep metaphysical sense of ``what there is''. When in fact we talk about a ``decoration of spacetime'', we are referring to a pattern of material stuff filling \emph{something}, namely, spacetime itself.\\
The above reasoning shows the need for integrating some (metaphysical and physical) considerations on space and time in a wider reflection on the status of primitive ontology approaches to quantum physics. This need becomes all the more compelling when inquiring into whether such a class of approaches can do a good job in mitigating or solving the conceptual problems that accompany the quest for a quantum theory of gravity. In canonical quantum gravity, for example, it is also the spacetime of general relativity to be quantized together with material fields, thus making impossible to consider it just as an inert arena where quantum phenomena take place. Very simply speaking, the absence of such an arena is referred to as \emph{background independence}.\\
Since the road to quantum gravity - and to a primitive ontology approach to quantum gravity - is still long and largely unexplored, here we will maintain a very modest attitude, and we will just start to investigate and assess the possibility of constructing a non-relativistic Bohmian theory of point-like particles where ``the stuff that is guided'' is in fact ``all there is''. The most simple way to do this is to show that equations (\ref{BM}) do not refer to external spatial and temporal structures that exist independently of the particle configuration but, instead, refer to a system of relations instantiated by particles. Of course, we could just argue in favor of a relationalist interpretation of (\ref{BM}) simpliciter - for example, following the steps of the remarkable work made in \cite{422} - but that would be an exercise in metaphysics not likely to have useful consequences in the quest for a Bohmian theory of quantum gravity. Suffices it to say that it is not at all guaranteed that such a searched-for relational interpretation would carry over to the case of fields. Having (also) this worry in mind, we will suggest a formal implementation of a relationalist-like approach to BM which could in principle be extended to the quantum gravitational regime.\\
So far, the most promising attempts to construct a purely relational version of a (classical) theory describing moving point-like particles were carried out by Julian Barbour and his collaborators.\footnote{In particular, Edward Anderson has later refined and expanded Barbour's relational framework while developing a new quantum gravity program. Cf. \cite{443} for his monumental review, which presents the state-of-the-art in relational mechanics and describes its developments since the inception of the first models.} In the following, we will first review motivations and results of Barbour's program, we will then investigate the possibility of applying this program to standard BM, and we will then inquire into the metaphysical consequences that a purely relational Bohmian theory would entail.

\section{Barbour's Program}
\subsection{Prolegomena}
In Newtonian mechanics (NM), the dynamics of a monogenic system\footnote{That is, a system subjected to a scalar potential that can be a function of coordinates, velocities, and time.} can be formulated as an action principle over the space $\mathcal{Q}\times\mathbb{R}$. $\mathcal{Q}$ is the configuration space, while $\mathbb{R}$ models the absolute Newtonian time.  The description of the physical system at each instant is given by a point $\mathbf{q}\in \mathcal{Q}$. Its evolution is given by its trajectory $\mathbf{q} =\mathbf{q}(t)$, which is a function of the time variable $t\in\mathbb{R}$, and which is found by solving the Euler-Lagrange equations, which follow from the so-called \emph{Hamilton's principle}. The principle holds that the actual path of the system between two fixed end points $(t_{1},t_{2})$ is the one that renders stationary the action $I$, which depends on the \emph{Lagrangian} function $L\equiv L(\mathbf{q},\frac{d\mathbf{q}}{dt},t)$:
\begin{equation}\label{lag}
\delta I=0,\quad I=\int_{t_{1}}^{t_{2}} Ldt.\footnote{Usually, the Lagrangian is taken to be the difference $T-V$ between the kinetic and the potential energies of the system. However, as we will see later, other choices are possible.}
\end{equation}
In the case of a system of $N$-particles we have that $\mathcal{Q}=\mathbb{R}^{3N}$, and the system is described by $\mathbf{Q}=\{\mathbf{q}_{1},\dots,\mathbf{q}_{N}\}$, $\mathbf{q}_{k}=(x_{k},y_{k},z_{k})$ being the position of the $k$-th particle in Euclidean $3$-space at a given time relative to a given coordinate system. Hence, we immediately notice that both NM and BM are formulated over the same configuration space, although their dynamical laws are obviously radically different.\footnote{It is worth noting that standard quantum mechanics is already formulated over the configuration space but only in BM do we see clearly the significance of this fact. This is because in BM, quantum dynamics is presented in such a way that the parallels between quantum and classical dynamics can be seen clearly. Of course, this is due to the fact that BM can be interpreted using beables only, which makes it possible to establish a fundamental continuity between quantum and classical dynamics on the meaning of dynamics - i.e. both are dynamics of point-like particles. The difference between them thus lies only in the form of the law of dynamics.}\\
The best-matching framework is an attempt to formulate a ``perfect'' theory of dynamics.\footnote{The language of ``perfect'' dynamics is due to the authors and not to the originators of the theory. In our view, this language is helpful to bring out the fundamental physical motivation underlying Barbour's program, namely, that it is looking for some criteria to evaluate what constitutes a ``perfect'' dynamics.} This begs the question: what constitutes a ``perfect'' formulation of dynamics? In other words, what are the minimum theoretical requirements for a theory of dynamics to be considered good? One possible answer is to say that a theory of dynamics must be formulated using only variables that are physically observable. One could think of this as an application of Occam's razor, namely, that the ``perfect'' formulation of dynamics should make use of a minimally necessary set of quantities (i.e. all the physically observable data only). We can call this the \emph{minimalist} requirement for ``perfect'' dynamics. One could argue that this requirement leads to a strictly relational dynamics since, empirically, we only observe angles and ratios of relative distances\footnote{We observe ratios of relative distances, not relative distances themselves since in measuring an object with a ruler, we are really comparing it to the ruler.} Indeed, this is the line of interpretation taken by Barbour and his collaborators.\footnote{See \cite{390,425}, for a technical introduction to Barbour's program, \cite{77} for an extensive survey of the philosophical motivations behind the program, and \cite{444} for a conceptual and technical expansion of the program.}\\
However, the minimalist requirement is not enough to secure a ``perfect'' theory of dynamics. This is because one could imagine formulating a useless theory using only variables that are physically observable but has no predictive power. It is necessary for any good theory of dynamics to have some degree of power for empirical prediction. A ``perfect'' theory should by definition satisfy this requirement in the best way. One way of implementing this requirement is to say that a ``perfect'' theory should be \emph{maximally predictive}, that is, it should be able to predict all subsequent motions using only the initial data that are physically observable.\footnote{This requirement is what Barbour calls Poincar\'e's principle. For a formulation of this principle, see \citet[][p. 302]{83}.} A ``perfect'' theory of dynamics thus has a minimalist and a maximalist requirement.\footnote{The shrewd reader might already object that no ``perfect'' universal classical theory of moving point-like particles can recover the full empirical predictions of Newtonian mechanics, for that would mean specifying among the initial data the rate of change (in absolute time) of the orientation of the overall configuration of particles with respect to absolute space. That would obviously violate the minimalist requirement. We will discuss this point later.\label{J}}\\
For our purposes, it is important to point out two key premises implicit in our discussion so far.  First, a philosophical notion of ``physical observability'' is presupposed in the above approach to understand the minimalist requirement. Hence it is assumed that it is possible to differentiate what counts as physically observable and what does not. Second, granted that if such a notion of physical observability is available, by setting this as a criterion for a ``perfect'' theory of dynamics, one assumes that quantities that are physically unobservable are redundant theoretical structures that are undesirable for theories of dynamics.\footnote{It is not obvious why this is the case. For instance, the wave function is not straightforwardly physically observable but according to the most common understanding of standard quantum mechanics, it is an essential theoretical structure for a successful formulation of the theory.} These premises are not straightforward and it seems that for anyone who wishes to motivate the best-matching framework in the above manner, she is forced to commit to these premises. While we are sympathetic to some form of the second premise due to our interest in implementing the requirement of background independence,\footnote{See \cite{433} for a preliminary discussion of the compatibility between Bohmian dynamics and the requirement of background independence.} the first premise is particularly difficult for our purposes here because any attempt to construct a background independent Bohmian theory cannot be smoothly grounded on notions such as physical observability, since particles' trajectories in BM are not observable in the same sense that Newtonian trajectories are.\footnote{This point is nicely illustrated in \cite{426}.} Thus it is not clear whether, philosophically, it is ever possible to implement the technical procedure of best-matching to BM without being inconsistent with its initial motivations.\\
Fortunately, Barbour also offers another way of interpreting the minimalist requirement. This is the idea that a ``perfect'' theory of dynamics should explain all of physics by geometry, in a purely Cartesian spirit.\footnote{See \citet[][especially section 1]{419} for this line of interpretation and the historical details.} We shall call it the \emph{geometrization of physics}. In the concrete case at hand, the geometrization program would amount to reducing (in a strong formal sense) the physical description of the particles' behavior - hence, the dynamics in primis - to geometrical features of a properly constructed fundamental space. We will be more explicit on this point in a moment. For now, we just remark that this interpretation of the minimalist requirement replaces the ``purely observational'' interpretation by demanding a ``perfect'' theory of dynamics to consist of only minimally necessary geometrical structures. A dynamics of this sort would be a theory where a minimally necessary amount of geometrical structures is utilized so that it can be maximally predictive. This alternative formulation of the minimalist requirement allows the Bohmian theorist to get out of the problem of physical observability. It is perfectly consistent to commit to Bohmian particles and to require a Bohmian theory to be maximally predictive with respect to its geometrical structures. What follows will largely be grounded on this notion of ``perfect'' dynamics.\\
Before we venture into the technical construction of the scheme, one last clarification is necessary. While it is perfectly legitimate to pursue a ``perfect'' dynamical formulation of BM according to the afore-mentioned principles, there is a much more direct motivation to adapt the above considerations. This is the fact that the application of the minimalist requirement in best-matching leads to a straightforward elimination of absolute space and time in dynamics. Thus if one finds the search for a ``perfect'' formulation of dynamics appealing, indeed one can adapt the above motivation for applying best-matching to BM. Nevertheless, a pragmatist could still find the approach of this paper useful in that ultimately it leads to the elimination of absolute space and time in BM.

\subsection{Implementing the Minimalist and the Maximalist Requirements}\label{maxmin}
Let us stick for the moment to the classical picture and concretely see how we can implement a ``perfect'' dynamics. The first step is to set up a geometrical arena that implements the minimalist requirement. To do so, we start from the configuration space $\mathcal{Q}$. $\mathcal{Q}$ contains certain geometrical symmetries specifiable by a Lie group $\mathcal{G}$ (for instance, the Euclidean group in Cartesian 3-space $\mathbb{R}^3$). However, in $\mathcal{Q}$, two configurations $\mathbf{Q}_1$ and $\mathbf{Q}_2$ are distinct even if one can be generated from the other purely by the action of $\mathcal{G}$. To fix the ideas, let us consider Euclidean 3-space and fix $\mathbf{Q}_1$ to encode the spatial coordinates of three particles mutually arranged in the shape of an equilateral triangle. If we take $\mathbf{Q}_2$ to be the result of applying a rigid translation to $\mathbf{Q}_1$ we immediately understand that $\mathbf{Q}_1$ and $\mathbf{Q}_2$ represent two different embeddings of the same relational configuration in Euclidean space. This is the sense in which $\mathcal{Q}$ does not satisfy the minimalist requirement: it features configurations which are relationally indistinguishable but still distinct with respect to some action of $\mathcal{G}$. The proper ``minimalist'' geometrical arena, call it $\mathcal{Q}_0$, would then arise by ``quotienting out'' the orbits of $\mathcal{G}$ from $\mathcal{Q}$. The metaphysical commitment behind this reasoning is crystal clear: there is nothing over and above particles and the mutual space-like relations they stand in that identifies a universal configuration. Therefore, any description that adds geometrical information to the relational one, e.g. how a configuration is embedded in Euclidean $3$-space, has to be taken as introducing irrelevant extra structure to the picture. Metaphysically speaking, this stance amounts to adopting a rather strong criterion of indiscernibility for configurations based on their shape alone.\\
To see more concretely how this metaphysical attitude is implemented in the theory, if we start from a $3N$ dimensional configuration space $\mathcal{Q}=\mathbb{R}^{3N}$ - where $N \geq3 $ is the number of point particles in the system - we can arrive at $\mathcal{Q}_0$ by successively quotienting $\mathcal{Q}$ into the orbits of the action of the group that consists of translations, rotations and dilations (i.e. homogeneous scalings). First, consider the translations. We can assign all configurations in $\mathcal{Q}$ that are carried into each other by Euclidean translations $\mathbf{r} \in \mathbb{R}^{3}$ to a common group orbit of $\mathbb{R}^{3}$, thus decomposing $\mathcal{Q}$ into the group orbits of $\mathbb{R}^{3}$. Similarly, we can carry out the same quotienting process for rotations $\mathbf{s}\in SO(3)$ and dilations $\mathbf{k}\in\mathcal{H}$, this latter group being roughly the group of Euclidean dilations or homothety-translations;\footnote{To be precise, uniform scalings are particular cases of homothety transformations. However, this level of precision is not essential for our purposes.} the group encompassing all the above transformations is referred to as the \emph{similarity group} $Sim(3)$ of Euclidean $3$-space. In general, the quotienting out operation is highly non-trivial, and gives rise to a reduced configuration space $\mathcal{Q}_0=\mathcal{Q}/Sim(3)$ that is a \emph{stratified manifold}, which, roughly, can be conceived as a ``union'' of manifolds of (possibly) different dimensions (the \emph{strata}).\footnote{See \citet[][especially section 2 and appendix B]{446}, for a self-contained technical discussion of the topic, including field theories. That article makes also clear that the stratified structure of $\mathcal{Q}_0$ carries physical import, so it should be accepted as an unavoidable element of (scale-free) relational theories.} We will see later how the non-trivial structure of $\mathcal{Q}_0$ forces some caveats upon the presented framework; for the time being we notice that this reduced configuration space represents the geometrical arena that satisfies a minimalist requirement faithful to the commitment to same-shape indiscernibility. We will call it, following Kendall's extensive work on the subject \citep{449},  \emph{shape space}.\\
The second step is to formulate a predictive dynamics on $\mathcal{Q}_0$.  The standard Newtonian dynamics relies on the notion of absolute time. In that case, dynamics is then formed in $\mathcal{Q}\times\mathbb{R}$. However, the postulate of a primitive, fixed ``temporal'' structure modeled by $\mathbb{R}$ once again violates the minimalist requirement if it is possible to formulate dynamical theories without it.\footnote{This remark makes manifest the fact that the pursuit of a ``perfect'' dynamics amounts to an implementation of spatial \emph{and} temporal relationalism.} Following Lanczos, Barbour realizes that it is indeed possible to formulate dynamical theories without time using Jacobi's principle.\\
Put it simply, Jacobi's principle \footnote{See \citet[][pp. 132-140]{414} for a short introduction.} is a way of formulating dynamical theories using a timeless\footnote{In the remainder of this section, we will refer to a ``timeless'' dynamics in the weak sense of ``without absolute time''. We will postpone to section \ref{metrbm} (in particular \ref{tfq}) the discussion of whether this kind of dynamics should be interpreted as giving up time entirely or still retaining some (weak) temporal structure.} variational principle that is parametrized by a $\lambda\in\mathbb{R}$:
\begin{equation}\label{jac}
\delta\mathcal{S}_J=0,\quad \mathcal{S}_J=2\int_{\lambda_{1}}^{\lambda_{2}} \sqrt {E-V}\sqrt {T_J} d\lambda,
\end{equation}
where $T_J = \frac{1}{2}\sum_{i=1}^{N} m_i\frac{d\mathbf{q}_i}{d\lambda}\frac{d\mathbf{q}_i}{d\lambda}$ is the parametrized kinetic energy, $V$ is the potential to which the particles are subjected, and $E$ is the total energy of the system.\\
There are two elements to notice about Jacobi's principle, which are crucial to our treatment. First, unlike the ``$t$'' variable in the standard variational formulation (\ref{lag}), here the parameter $\lambda$ is entirely arbitrary since $S_J$ is invariant under arbitrary transformations of the form $\lambda \rightarrow f(\lambda)$. The only requirement is that $\lambda$ is monotonically increasing, since it has to act as a ``time-like'' label. Second, (\ref{jac}) makes manifest the geometrization of the problem of finding the appropriate dynamics of a N-particle system. If, in fact, we drop the perspective of $\mathcal{Q}$ being a flat space whose line element is the Euclidean one $ds^{2}=\sum\delta_{ij}d\mathbf{q}_{i}d\mathbf{q}_{j}$, and we adopt a new non-trivial Riemannian structure for it given by $ds^{2}=2(E-V)T_{J}d\lambda=(E-V)\sum m_{i}\delta_{ij}d\mathbf{q}_{i}d\mathbf{q}_{j}$, then we immediately realize that (\ref{jac}) is nothing but a geodesic principle on this curved version of $\mathcal{Q}$.\footnote{\label{confr}In other words, $(E-V)$ plays the role of a conformal factor. This of course means that, in order for the formalism to make sense, such a factor should be well-behaved enough (e.g. no zeros, infinities, or non-smothnesses).} The determination of the system's dynamics is thus reduced to a purely geometrical problem, i.e. the determination of  the geodesics of the curved configuration space. This point gives rise to an interpretation of Jacobi's principle more commonly discussed in the literature as the geometrization of mechanics,\footnote{See, for example, \citet[][chapter I section 5, chapter V section 7, and chapter VIII section 9]{414}.} which provides an insightful way to see the conceptual continuity between classical mechanics and Einstein's theory of general relativity.  As we will see in the next section, what is significant about Barbour's use of Jacobi's principle lies in his interpretation of the principle within the conceptual framework of best-matching.\\
What is the relation between the just sketched timeless dynamics and the standard Newtonian one? As a matter of fact, one can actually easily recover the Newtonian formulation from the above dynamical framework. To see this, consider that the Jacobi's principle is formally equivalent to (\ref{lag}) if we put:
\begin{equation}
\mathcal{L} =2\sqrt {E-V}\sqrt {T_{J}}.
\end{equation}
By performing the variation with respect to the configuration variables, we get the appropriate Euler-Lagrange equations:
\begin{equation}\label{eulag}
\frac{d}{d\lambda}\left(\frac{\partial \mathcal{L}}{\partial (d\mathbf{q}_i/d\lambda)}\right)=\frac{\partial \mathcal{L}}{\partial \mathbf{q}_i}\quad\Rightarrow\quad\frac{d}{d\lambda}\left(m_i \sqrt {\frac{E-V}{T_J}} \frac{d\mathbf{q}_i}{d\lambda}\right)=-\sqrt {\frac{T_J}{E-V}}\frac{\partial V}{\partial \mathbf{q}_i}.
\end{equation}
(\ref{eulag}) fixes the structure of canonical momenta:
\begin{subequations}\label{moms}
\begin{equation}\label{moms1}
\mathbf{p}_i = \frac{\partial \mathcal{L}}{\partial (d\mathbf{q}_i/d\lambda)} = m_i \sqrt {\frac{E-V}{T_J}} \frac{d\mathbf{q}_i}{d\lambda};
\end{equation}
\begin{equation}
\frac{d\mathbf{p}_i}{d\lambda} =\frac{\partial \mathcal{L}}{\partial \mathbf{q}_i}= -\sqrt {\frac{T_J}{E-V}}\frac{\partial V}{\partial \mathbf{q}_i}.
\end{equation}
\end{subequations}
Recall that in this formulation of dynamics, $\lambda$ is an arbitrary parameter. Now if we choose $\lambda$ such that:
\begin{equation}\label{lt}
\frac{T_J}{E-V}=1\Rightarrow E=T_{J}+V,
\end{equation}
and we substitute it in (\ref{moms}), then we recover NM with respect to this particular value of the parameter $\lambda$, call it $t$:
\begin{equation}\label{yal}
\mathbf{p}_i =  m_i \frac{d\mathbf{q}_i}{dt}, \hspace{1cm} \frac{d\mathbf{p}_i}{dt} = -\frac{\partial V}{\partial \mathbf{q}_i}.
\end{equation}
Let us carefully reflect on the afore-mentioned framework. Recall that we are considering a universal theory of particles. According to the ``orthodox view'' that takes seriously the commitment to absolute time, condition (\ref{lt}) is not a mere choice, but a pre-existing condition (energy conservation). Consequently, the Jacobi's principle (\ref{jac}) is just a different way to put the Hamilton's principle (\ref{lag}), this latter being the real fundamental principle behind classical dynamics. On Barbour's view, the tables are instead turned. If we endorse the minimalist requirement for a ``perfect'' dynamics, we cannot but consider (\ref{jac}) as the fundamental principle of dynamics. From this point of view, (\ref{lt}) is indeed a mere choice akin to a gauge fixing: NM can be thought of as a particular choice of parametrized dynamics that gives rise to the simplest form for the equations of motion. Put it plainly, the relation between timeless and Newtonian dynamics considered under the light of the minimalist requirement shows that absolute time can be seen as an emergent quantity in the context of (particle) dynamics. As Butterfield puts it:
\begin{quote}
[B]arbour provides examples of theories in which a temporal [...] metric is emergent in the strong sense of being fully definable from the rest of the physical theory. So this is emergence in as strong a sense of reduction as you might want.\\
\citep[][p. 296]{402}
\end{quote}
One of the main reasons for adopting a universal perspective in this framework is now manifest. If, in fact, we consider only subsystems of particles, we could apply the Jacobi's principle to each of them separately, obtaining the same results shown above. However, when coming to fixing the parameter leading to the simplified equations of motion (\ref{yal}), each system would have its own condition (\ref{lt}). As a consequence, each subsystem would have its own clock and, in general, all these ``local'' clocks would not march in step. Hence, condition (\ref{lt}) applied to the universe as a whole is the only option that assures that all the subsystems' clocks will share the same absolute time $t$.\footnote{See \citet[][section 9.6]{443} for a critique of the ``marching in step'' criterion.}\\
We have then achieved a remarkable result, but this is not yet the entire story. Although Jacobi's principle as formulated above represents a huge leap into the implementation of a minimalist dynamics, still it is not enough for our purposes, at least as it stands right now. This conclusion is quite obvious, since (\ref{jac}) will get us to a timeless dynamics on $\mathcal{Q}$, not on $\mathcal{Q}_0$ (which is what we need). But is it possible to use a Jacobi-like principle on $\mathcal{Q}_0$ to formulate a predictive dynamics? We will now turn to this problem.

\subsection{The Best-Matching Procedure: Timeless Dynamics on $\mathcal{Q}_0$}\label{besme}
The most natural approach is to attempt to write down a Jacobi-like geodesic principle on $\mathcal{Q}_0$ directly based on the ``shape'' variables $q_0 \in \mathcal{Q}_0$. However, it turns out that this is extremely difficult.\footnote{One could attempt to formulate such a dynamics via relative coordinates such as inter-particle distances $r_{ij} = |\mathbf{q}_i - \mathbf{q}_j|$. However, such theories notoriously suffer from predicting anisotropic masses that disagree with empirical observations. See \citet[][especially section 6]{421} for a discussion and references.} A more practical approach is (i) to define a metric $T_{JBB}$ that measures how much two relational configurations differ, and (ii) to construct such a metric using variables defined on $\mathcal{Q}$: this approach was first taken by \citet{83} and later refined by Anderson.\footnote{See \citet[][sections 1.5-1.8, 2.A]{443} for a technical overview on the evolution of the formalism (from the original one in \citealp{83}, which spoiled the reparametrization invariance of the action, to the one presented here, which fixed the issue), and \cite{455} for a detailed technical discussion.} The following presentation will be mainly based on \citet[][sections 1-2]{454}, \citet[][example 4]{447}, and \citet[][sections 1-4]{419}.\\
Consider two distinct N-particle relational configurations $q_0^1, q_0^2 \in \mathcal{Q}_0$. We can coordinatize them by two Cartesian coordinate frames. Now the first frame can be laid down in an entirely arbitrary way relative to $q_0^1$ (reflecting, e.g., the freedom to choose the origin of the frame). To define a ``distance'' $\mathcal{D}$ between $q_0^1$ and $q_0^2$, we require that the second frame is laid down relative to $q_0^2$ using infinitesimal $\mathcal{G}$-transformations ($\mathcal{G}=Sim(3)$ as discussed in \ref{maxmin}) such that $\boldsymbol\updelta \mathcal{D}(\delta\mathbf{q}_i) = 0$. $\delta\mathbf{q}_i$ is the ``intrinsic'' infinitesimal overlap deficit between the two configurations; it is given in terms of the Cartesian coordinates $\mathbf{q}_i$ plus ``$\mathcal{G}$-frame'' corrections. More precisely, we have:
\begin{displaymath}
\delta\mathbf{q}_{i}=d\mathbf{q}_{i}-d\mathbf{A}-d\mathbf{B}\times\mathbf{q}_{i}+dC\cdot\mathbf{q}_{i},\footnote{Besides the already mentioned \cite{454, 447}, see also \citet[][sections 2.2.1, 2.3.1, 2.3.2 ]{443} for a discussion of this kind of ``differential'' and its relation to Lie derivatives.}
\end{displaymath}
where $\mathbf{A}$ and $\mathbf{B}$ are $3$-vectors accounting, respectively, for translations and rotations, while $C$ is an appropriate scaling function.
This procedure is aptly called \emph{best-matching} because the ``distance'' between two configurations in $\mathcal{Q}_0$ is defined by best-matching the two coordinate frames with respect to each other.\footnote{More pictorially, carrying out the procedure for which $\boldsymbol\updelta \mathcal{D}(\delta\mathbf{q}_i) = 0$ amounts to ``juxtapose'' $q_0^2$ with $q_0^1$ such that they fit in the best way possible. Clearly, in the limit case where $q_0^1$ and $q_0^2$ represent the same shape, this ``juxtaposition'' will make them perfectly overlap: their ``distance'' is then zero.}\\
The metric $T_{JBB}$ (where ``JBB'' stands for ``Jacobi-Barbour-Bertotti'') that implements the best-matching procedure by using calculational tools defined on $\mathcal{Q}$ (or, better, on $\mathcal{Q}\times\mathcal{G}$) is simply:
\begin{equation}\label{TJB}
\begin{split}
T_{JBB}=\frac{1}{2} \sum_{i=1}^N m_i \left(\frac{d\mathbf{q}_i}{d\lambda}- \frac{d\mathbf A}{d\lambda}-\frac{d\mathbf B}{d\lambda}\times\mathbf q_i+\frac{d C}{d\lambda}\cdot\mathbf q_i\right) \cdot \\ \cdot \left(\frac{d\mathbf{q}_i}{d\lambda}- \frac{d\mathbf A}{d\lambda}-\frac{d\mathbf B}{d\lambda}\times\mathbf q_i+\frac{d C}{d\lambda}\cdot\mathbf q_i\right),
\end{split}
\end{equation}
which is parametrized by the usual $\lambda$.\\
In order to determine the dynamics on $\mathcal{Q}_0$, first, we write down the following variational principle:
\begin{equation}\label{jacbar}
\delta \mathcal{S}_{JBB} = 0, \hspace{1cm}  \mathcal{S}_{JBB}= \int \sqrt {E-V}\sqrt {T_{JBB}}d\lambda.\footnote{$T_{JBB}$ is homogeneous of degree $2$, while E-V must be homogeneous of degree $-2$ (see \eqref{3}). As Anderson notes \cite[][section 2.3.2]{443}, since we are dealing with a scale-invariant theory, it would be far more geometrically natural to render both terms homogeneous of degree $0$ by, respectively, dividing and multiplying by the total moment of inertia $I$. However, as Anderson himself acknowledges, the form \eqref{jacbar} is the mechanically-natural one, since $T_{JBB}, E$, and $V$ bear the usual physical units ($I$ acting in this context just as a constant ``conversion factor'' between the two formulations). We then prefer to stick to this latter representation, which will make clearer the extension of the framework to BM discussed in section \ref{rbm}.}
\end{equation}
Then we perform a (free end point) variation with respect to the $\mathcal{G}$-auxiliaries $\mathbf A , \mathbf B , C$.\footnote{See \cite{456,453} for two fully worked-out models involving this formalism. Earlier work on ``Barbour-Bertotti'' models include, notably, \cite{462,463}.} This procedure gives rise to the following four constraints.
First of all, we note that the best-matched momenta have the same form of (\ref{moms1}), that is, they are ``direction cosines'' with respect to the kinetic metric $T_{JBB}$ multiplied by the term $(E-V)$. Hence, not all the momenta are independent (intuitively, we will have groups of momenta ``pointing in the same direction'' modulo a common factor), and this fact is encoded in the first constraint:
\begin{equation}\label{0}
\frac{1}{2} \sum_{i=1}^{N} \frac{\mathbf{p}_i\cdot \mathbf{p}_i}{m_i}-E+V=0.
\end{equation}
The second constraint arises from the variation of $\mathbf A$ It amounts to the vanishing of the total momentum of the system:\footnote{All the three relations (\ref{1}), (\ref{2}), (\ref{3}), hold in the center-of-mass frame.}
\begin{equation}\label{1}
\mathbf{P}= \sum_{i=1}^{N} \mathbf{p}_i = 0.
\end{equation}
This condition implies that the system is isolated, which is consistent with the universal perspective adopted in the framework. Such a condition is propagated if the potential $V$ is invariant under spatial translations.\\
The third constraint answers the objection briefly considered in footnote \ref{J}. It is a consequence of $\mathbf B$-variation and states that the total angular momentum of the system vanishes:
\begin{equation}\label{2}
\mathbf{J} = \sum_{i=1}^{N} \mathbf{q}_i \times \mathbf{p}_i = 0.
\end{equation}
(\ref{2}) explains why the present framework is maximally predictive even if the initial data required do not include any information regarding the orientation of the system in absolute space. In this case, to secure the propagation of the constraint, the potential $V$ has to be invariant under rotations, which is the case if it is a function of the interparticle separations.\\
The fourth constraint (arising from $C$-variation) gives rise to the most unintuitive consequences. It amounts to the vanishing of a quantity that, in analogy with the former designations, we can call \emph{dilational momentum}:
\begin{equation}\label{3}
\mathbf{D} = \sum_{i=1}^{N} \mathbf{p}_i \cdot \mathbf{q}_i = 0.
\end{equation}
Condition (\ref{3}) is consistent with the framework if $V$ is homogeneous of degree $-2$ in the positional variables \emph{and} the total energy $E$ vanishes in the ``Newton'' gauge.\footnote{The concrete calculations are carried out in \citet[][section 2]{419}.} The consequence of these restrictions is, indeed, remarkable: motions for which $V=const.$ and $E>0$ - that is, inertial motions - are not allowed. Barbour stresses this fact as follows:
\begin{quote}
It is in this sense that inertia violates scaling. \emph{There is no maximally predictive inertial dynamics on shape space}. One cannot formulate a theory of pure inertial motion without introducing additional kinematic structure - an absolute scale of length - that mathematical intuition suggests one should not employ. [I]f one wishes to have any dynamics at all on shape space that satisfies the Poincar\'e criterion, \emph{it must include forces and have vanishing energy}.\\
\citep[][p. 1546, Barbour's emphases]{419}
\end{quote} 
We will see later how the very last sentence in the above quotation can acquire a new sense in a Bohmian framework.\\ 
In the end, by solving the constraints, one can eliminate the $\mathcal{G}$-auxiliaries from the action \eqref{jacbar}, thus finding the ``real'' geodesic principle on $\mathcal{Q}_{0}$. We are now in the position of applying the same exact reasoning behind Jacobi's principle that led from (\ref{jac}) to (\ref{eulag}). The final result consists of equations of motion of the same form as (\ref{eulag}). Also in this case, by fixing the ``Newton'' gauge through a condition analogous to (\ref{lt}), we get the usual Newtonian equations (\ref{yal}). Therefore, in the case of a timeless dynamics in shape space, we have that both spatial and temporal degrees of freedom of NM are recovered by applying the variational principle (\ref{jacbar}) and then fixing the ``Newton'' gauge $\lambda\rightarrow t$. This is the strong formal sense in which Newtonian space and time are reduced to the geometrical features of $\mathcal{Q}_{0}$.\\
However, some technical caveats are required at this point. The first is that no usual Newtonian potential is compatible with (\ref{3}), since normally - e.g. in the gravitational case - they are homogeneous of degree minus one. This is not by itself an insurmountable problem since, as discussed in \citet[][section 5.1.2]{443}, it is always possible to find some mathematical trick that mimics the form of the most usual classical potentials. However, this sort of trickery might lead to unwanted physical restrictions, such as no angular momentum exchange between subsystems.
This point is of course delicate, and we will see that the Bohmian context fully inherits this conceptual problem.\\
The second caveat regards the implementation of a geodesic principle on a general shape space. Given, in fact, that the global geometry of such a space is that of a stratified manifold, it is problematic to rigorously account for a dynamical evolution whose related geodesic trajectory hits different strata of $\mathcal{Q}_{0}$ (see \citealp[][section 9.4 and references mentioned therein]{446}, for discussion). This means that the above described framework works well in a suitably small region of $\mathcal{Q}_{0}$, but might break down on a larger scale, depending on the particular geometrical structure of $\mathcal{Q}_{0}$.\\
To recap: the best-matching framework in the version presented here represents an attempt at reducing Newtonian dynamics to a more fundamental theory where absolute space and time play no relevant role. The sense of reduction intended is very strong, i.e. a purely formal one: best-matching treats all the degrees of freedom which are not intrinsic to a universal configuration of particles (i.e. that are not given in terms of ratios of distances and angles), hence in primis spatial and temporal degrees of freedom, as mere gauge. This separation of degrees of freedom into physical and gauge is justified by the constraints (\ref{0}), (\ref{1}), (\ref{2}), and (\ref{3}) arising from the implementation of the Jacobi-Barbour-Bertotti's principle on shape space. As a result the $J=0$ subsector of NM is recovered\footnote{We note that nothing speaks against the possibility of constructing a classical theory that accounts for global $J\neq0$ effects in terms of change in the spatial relations of particles only. Of course, such a theory would exhibit a mathematical structure far more complicated than the present one.} from the underlying best-matching theory via a gauge fixing that sets a privileged temporal metric and a privileged length scale, from which inertial motions can be recovered (at least in sufficiently small regions of shape space). This also explains in what sense the best-matching dynamics can be considered as maximally predictive. Moreover, the fact that the dynamics is implemented as a geodesic principle over a curved Riemannian space (viz. shape space) represents a genuine reduction of all the salient dynamical features to geometrical ones: in this sense, the theory exhibits a dynamics that satisfies the minimalist requirement.

\subsection{Quantization}\label{QA}
There are currently a number of quantization procedures proposed for relational models, which, in some cases, were completed in detail (see \citealp[][sections 13-16]{443} for extensive discussion and detailed calculations):\linebreak \citet{457, 458,459} are notable examples. However, none of these - to our knowledge - were carried out in a primitive ontology framework (one exception being \citealp[][appendix A]{461}).\\
Here we just note that, if a ``perfect'' dynamics is a timeless theory or, better, a reparametrization invariant theory, then the canonical quantization of such a theory will give rise to a static universal wave equation.\footnote{See \cite{135}, \citet[][sections 3.1 and 3.4]{142}. However, as we will see in section \ref{metrbm}, not everybody agrees on this.}  To see this, it is sufficient to note that the canonical procedure for quantizing the theory will comprise the implementation of the constraints (\ref{0}), (\ref{1}), (\ref{2}), and (\ref{3}) as restrictions over the physically allowed wave functions.\footnote{Unless they are solved prior to quantization, which would lead to the same result considered above.} For example, the canonical quantization of the constraint (\ref{0}) will straightforwardly give rise to a time-independent Schr\"odinger equation:
\begin{equation}\label{pwdw}
\hat {\mathcal{H}}\Psi = E \Psi.
\end{equation}
In other words, the wave equation just takes definite values of $E$ for each configuration of the system. Imposing the other quantum constraints would further restrict the allowed universal wave functions to those that are eigenfunctions of the Hamiltonian operator with energy eigenvalue E=0. In the end - and not surprisingly - we see that there are compelling arguments in favor of the fact that a canonically quantized version of the theory considered above would exhibit a timeless dynamics dictated by a Wheeler-DeWitt-like equation:\footnote{For a detailed technical discussion of the quantization of theories that dispense with space and time as fundamental notions, see \cite{423}.}
\begin{equation}\label{wdw}
\hat {\mathcal{H}}\Psi = 0.
\end{equation}
As we shall argue, this outcome in actual fact unifies the perspective of Bohmian and Barbour's programs in a powerful way. This is because on the Bohmian view adopted here, that is the one that takes the wave function as a law-like element of the formalism,\footnote{A presentation and defense of this view can be found in \cite{428}.} a static wave equation from the universal perspective is exactly what one should expect.\\
Since the above overview have provided the reader with enough information on the best-matching procedure and its quantization, we can now turn to BM and see (i) how this framework might - or might not - apply in this context (the next section), and (ii) what are the possible metaphysical consequences of a fully worked out relational Bohmian theory of particles (section \ref{metrbm}).

\section{Relational Bohmian Mechanics: A Brief Sketch}\label{rbm}
If we agree that best-matching is a really promising framework in developing a truly relational mechanics - at least, but not only, for particle mechanics - it is then interesting to investigate whether pursuing a relational Bohmian particle theory might profit from adopting such a framework. Indeed the Barbour and Bohmian approaches share a common ``universal'' perspective: the aim of both approaches is to end up with a theory that describes the entire universe as a unique (``undivided'') system and then seek to recover the description of a subsystem of it as a suitable approximation of the behavior of this part with respect to the rest of the system. To be fair, however, the two approaches have to appeal to such a universal perspective for different reasons. In Barbour there is the need for a description of motions that is maximally predictive albeit disregarding a huge part of Newtonian initial data (especially angular velocity), while in Bohm there is the need to account for the appearance of a collapse of the wave function of a subsystem while retaining the globally non-local behavior of the system. It is remarkable that such original motivations are not only compatible, but even similar: both, in fact, are intended to account for the appearance of certain well-known features (e.g., absolute time, quantum collapses) at the level of subsystems while, in fact, denying the fundamental reality of such features. Having explained the motivation for pursuing a relational dynamics for Bohmian particles in the introduction, and having showed the virtues (and vices) of a relational mechanics based on best-matching in the previous section, we can now focus on a concrete proposal on how to implement a relational version of BM (RBM, for short) based on best-matching.\\
The most obvious strategy to create a relational quantum theory of $N$ point-like particles that satisfies the minimalist and the maximalist requirements would be to quantize the classical best-matched theory in the way suggested at the end of section \ref{QA}, thus ending up with a Wheeler-DeWitt-like equation on the relational configuration space $\mathcal{Q}_{0}$. At this point, in order to ``go Bohmian'', we should first of all realize that the theory whose dynamics is encoded in (\ref{wdw}) suffers from the standard conceptual problems arising both in quantum physics in general, and in particular in those theories constructed by canonically quantizing reparametrization invariant classical theories. Among these issues, we might mention the problem of making sense of superpositions or collapses of the universal wave function, and the problem of extracting a non-trivial dynamics from a  timeless equation of the form (\ref{wdw}). In this context, appealing to the insights that the Bohmian theory might give to this theory is a legitimate move even if, obviously, it is not the only one possible!\footnote{\cite{161} provide a clear review of the conceptual pros of going Bohmian in the context of a canonically quantized theory, especially in quantum canonical general relativity.}\\
After having argued that a Bohmian approach might be useful in this context, we could proceed by constructing a relational guiding equation for the $N$-particle system: this methodology would closely resemble the non-relational one adopted in \citet[][section 3]{222}, which consists in setting up a velocity field $\mathbf{v}^{\Psi}=\frac{d\mathbf{Q}}{dt}$ over $\mathbb{R}^{3N}$ depending on the wave function selected by the Schr\"odinger equation, which satisfies a number of symmetry conditions such as Galilean invariance and equivariance.\footnote{More precisely, the velocity field should be chosen such that the probability distribution $|\Psi|^{2}$ is equivariant with respect to it.} However, a moment of reflection shows that applying this strategy in the present case would not be so straightforward as it should prima facie seem. First of all, it would be practically impossible to work directly with the ``shape coordinates''  available in $\mathcal{Q}_{0}$ (see \citealp[][section 4]{419}, for a clear statement in this sense). Secondly an important amount of work should be done to show that (at least) a reparametrization invariant ``velocity'' field $\boldsymbol{\mathfrak{v}}^{\Psi_{0}}=\frac{\delta\boldsymbol{\mathfrak{Q}}}{\delta\lambda}$ can be defined over $\mathcal{Q}_{0}$, which is (i) compatible with the Wheeler-DeWitt-like equation (\ref{wdw}), and (ii) selects geodesic trajectories over $\mathcal{Q}_{0}$. Thirdly, it should have to be shown how from these two purely relational equations we could recover the standard Schr\"odinger equation plus the guiding equation of BM by fixing a particular value for the parameter $\lambda$. These remarks are not meant to suggest that pursuing this strategy would necessarily lead to no or wrong results, but to motivate the search for an alternative simpler methodology. In our opinion such a simpler strategy is available and can be arrived at by the following reasoning: since the subject matter of BM is basically the same of classical mechanics, namely the description of point-like particles moving in Euclidean $3$-space at an absolute time rate, and given that the dynamics of BM obeys the same dynamical symmetry conditions of classical mechanics, why don't we try to arrive at a version of RBM by straightforwardly applying best-matching to standard BM? This is the program that we are going to sketch in the remainder of this section.\\
In order to pursue our strategy, we need first of all to bring about as much as possible the similarities between classical mechanics and BM. We start, then, by reformulating (\ref{BM}) in a different albeit equivalent manner, namely, the way it was originally proposed in \cite{187a,187b} and thoroughly developed, e.g., in \cite{360}. This is done by considering two real functions $R(\mathbf{Q},t)$ and $S(\mathbf{Q},t)$ over $\mathbb{R}^{3N}\times\mathbb{R}$ such that, given a solution $\Psi(\mathbf{Q},t)$ of (\ref{sch}), it is the case that $\Psi=Re^{iS}$.\footnote{Here we gloss over the boundary and continuity conditions that must be placed to ensure that $\Psi$ - and hence also $S$ and $R$ - is physically meaningful.} Substituting this latter form of $\Psi$ in (\ref{sch}) and separating the real and imaginary parts of the resulting formula, we end up with the following two coupled relations:
\begin{subequations}\label{BM2}
\begin{equation}\label{HJ}
\frac{\partial S}{\partial t}+\sum_{i=1}^{N}\frac{(\nabla_{i}S)^{2}}{2m_{i}}+V+\mathcal{V}=0,
\end{equation}
\begin{equation}\label{cons}
\frac{\partial R^{2}}{\partial t}+\sum_{i=1}^{N}\nabla_{i}\Bigg(R^{2}\frac{\nabla_{i}S}{m_{i}}\Bigg)=0.
\end{equation}
\end{subequations}
Since our starting theory is (\ref{BM}), that is a non-relativistic theory of $N$ point-like particles, what we have done is basically to rewrite it as a Hamilton-Jacobi theory. We see this by looking at (\ref{HJ}) and recognizing that it becomes the Hamilton-Jacobi equation of our system if we assume that the $k$-th particle velocity is $\frac{\nabla_{k}S}{m_{k}}$. Under this reading, (\ref{cons}) states the conservation of $R^{2}=|\Psi|^{2}$ along the particles' trajectories. Pursuing this classical analogy leads to the introduction of a further ``quantum'' potential of the form:
\begin{equation}\label{qp}
\mathcal{V}=-\sum_{i=1}^{N}\frac{1}{2m_{i}}\frac{\nabla_{i}^{2}R}{R}.
\end{equation}
The reason why we call it a potential becomes manifest if we derive (\ref{HJ}) with respect of $\boldsymbol{\nabla}_{k}$, thus arriving at a ``Newtonian-like'' equation of motion for the $k$-th particle, which reads:
\begin{equation}\label{qnw}
m_{k}\frac{d^{2}\mathbf{q}_{k}}{dt^{2}}=-\boldsymbol{\nabla}_{k}\big(V+\mathcal{V}\big).\footnote{It is possible to arrive at the very same expression by differentiating (\ref{gui}) with respect to time, which stresses the fact that the two formalisms are equivalent.}
\end{equation}
It is extremely important to clarify that our interest in adopting this Newtonian disguise for BM is purely formal and does not entail that we are committing us to things such as quantum forces exerted by some kind of $\Psi$-field. Our commitments remain firmly those compatible with the version of the theory given by (\ref{BM}), which in particular means that we are not reifying $\Psi$ in any way.\footnote{Otherwise, one of our key motivations for pursuing this program, i.e. finding a Bohmian theory of $N$-particles with a more parsimonious and coherent ontology than the standard one, would be betrayed.} However, the advantage of casting BM in a Newtonian form should be clear to the reader, since now we can repeat for this theory - mutatis mutandis - the same reasoning that, in the previous section, led to a best-matched version of classical mechanics. In the present case, the key physical quantities that enter best matching are the kinetic energy:
\begin{equation}\label{kine}
T=\frac{1}{2}\sum_{i=1}^{N}m_{i}\frac{d\mathbf{q}_{i}}{dt}\frac{d\mathbf{q}_{i}}{dt}=\sum_{i=1}^{N}\frac{(\nabla_{i}S)^{2}}{2m_{i}},
\end{equation}
and the total potential energy:
\begin{equation}\label{totpot}
\mathfrak{V}=V-\sum_{i=1}^{N}\frac{1}{2m_{i}}\frac{\nabla_{i}^{2}R}{R}.
\end{equation}
The first delicate point is to choose the appropriate set of symmetries with respect to which perform the best-matching procedure. Since both $T$ and $\mathfrak{V}$ are Galilean invariant,\footnote{See \citet[][section 3.11]{360} for a discussion of the invariance properties of the theory (\ref{BM2}).} we can best-match these quantities with respect to the same gauge group $Sim(3)$ used for classical mechanics (but nothing prevents us from considering a different or more extended gauge group, if needed). However, note that in this case, the kinetic energy (\ref{kine}) has in general a non-trivial form due to its dependence on the square of the spatial gradient of the wave function's phase $S$. This complication is needed in order to implement into the theory the requirement that the velocity of the particles depends on the spatial gradient of $S$ (otherwise equation (\ref{qnw}) would select a broader class of motions than those allowed by  (\ref{BM})). The first conceptual issue thus reads:
\begin{fed}\label{prob1}
Is it possible to construct a kinetic metric of the form (\ref{TJB}) from (\ref{kine})?
\end{fed}
A detailed answer to this question would require a technical paper on its own. Since our purpose here is, much more modestly, to show that RBM is not a priori impossible to construct, we will be content to point out a case in which the answer to the above question is positive: this very simply happens when the phase does not depend on time. Clearly, when the phase has this form, then the right hand side of (\ref{kine}) will not depend on time as well, and it would be easy to implement best-matching by putting the whole expression in the form (\ref{TJB}). Actually, one may have thought of a broader class of phases, namely those for which the positions and time dependencies are separable, that is, $S(\mathbf{Q},t)=S'(\mathbf{Q})-\mathcal{E}t$, with $\mathcal{E}$ the total energy of the system in the ``Newton'' gauge. However, this class of phases are not consistent with the requirement of the total energy of the system being zero in the ``Newton'' gauge.\\
The conceptual issue \ref{prob1} is not the only one we would face when trying to implement a ``Bohmian'' Jacobi-Barbour-Bertotti principle as in the classical case. The second issue, in fact, regards how to construct the conformal factor $\sqrt{\mathcal{E}-\mathfrak{V}}$ that ``bends'' the kinetic metric $T_{JBB}$, thus generating a non-trivial timeless dynamics on $\mathcal{Q}_{0}$ (modulo the caveats discussed at the end of section \ref{besme} and in footnote \ref{confr}):
\begin{fed}\label{prob2}
Is it possible to construct a conformal factor of the form $\sqrt{\mathcal{E}-\mathfrak{V}}$ in order to implement a geodesic principle resembling (\ref{jacbar})?
\end{fed}
Also in this case, the generic form of (\ref{totpot}) does not permit a quick answer. This is obviously because the additional quantum part (\ref{qp}) of the potential introduces a highly non-trivial dependence on the wave's amplitude $R(\mathbf{Q},t)$ and its second spatial derivatives. To solve this issue, two different routes can be taken. The first, and most perilous, is to modify the Jacobi principle as follows. Since this principle singles out the geodesics of the shape space $\mathcal{Q}_{0}$, intended as a curved Riemannian manifold, a possible strategy would be to include the ``quantum'' part of the potential as characterizing some non-trivial metrical property of this space other than its curvature: for example, its torsion. This strategy would require a relevant amount of work to be carried out, but nothing prevents it a priori from being successful. The second strategy would more simply amount to considering all the relevant cases and check by calculation if the Jacobi principle can be effectively implemented. Also in this case, considering amplitudes $R(\mathbf{Q},t)=R'(\mathbf{Q})$ that do not depend on time would do the job, at least when best-matching (\ref{totpot}) with respect to translations and rotations. However, when considering scale invariance, the problem becomes extremely delicate:
\begin{fed}\label{prob3}
Is it possible to construct a total potential $\mathfrak{W}\equiv\mathfrak{W}(\mathfrak{V})$ which is homogeneous of degree $-2$ but, at the same time, gives rise to an equation of motion of the form (\ref{qnw}) in an appropriate limit?
\end{fed}
This issue seems the most compelling among the three we pushed forward so far, but it is also the one more likely to bring new physics into the picture, independently of the final answer to the question. In case it turned out that this issue cannot be solved, then we should surrender to the fact that BM can be given at best only a ``mild'' relationalist implementation, where inertial effects cannot be fully reduced to geometric facts holding in shape space, but we would also gain some insight on the possible quantum aspects at the roots of inertia. On the other hand, if the question could be answered in the positive, then  the discrepancies between the ``real'' universal dynamics encoded in $\mathfrak{W}$ and the ``observed'' one encoded in $\mathfrak{V}$ would most likely be based on new testable physical assumptions about the universe.\\
To conclude this section, we should consider a fourth issue which, in some sense, summarizes the previous three:
\begin{fed}\label{prob4}
Assuming that a version of RBM is actually implementable, how much of BM could be recoverable from it?
\end{fed}
If RBM could be implemented, then according to the constraint (\ref{0}), and the condition that the total energy of the system should vanish in the ``Newton'' gauge, we would obtain a Hamilton-Jacobi equation (\ref{HJ}) of the form $\frac{\partial S}{\partial t}=0$. This would be entirely consistent with the simple solution of the conceptual issue \ref{prob1} proposed above. As regards the amplitude $R$, however, the constraints do not straightforwardly select any of its characteristic features\footnote{Nonetheless, $R$ does have a general distinctive feature that might be interesting in this context, that is, the fact that it influences the form of the quantum potential (\ref{qp}) modulo a multiplicative constant (i.e. $\mathcal{V}$ does not change under transformations $R\rightarrow kR, k\in\mathbb{R}$). This means that the physical information encoded in $R$ which determines (\ref{qp}) is insensitive to scaling transformations.} (which enforces the considerations made about the conceptual issue \ref{prob2}). Anyway, if we expect RBM to be consistent with the ``timelessness'' of a reparametrization invariant quantum theory, then it is likely that also $R$ would turn out to be a function independent of time. In other words, RBM would recover the sector of BM with stationary (superpositions of) wave functions corresponding to zero total energy of the system. The fact that the standard best-matching framework would not recover the full content of BM is not by itself a problem. Even the best-matched theory based on (\ref{jacbar}) was not able to recover the full Newtonian dynamics but this is not an issue insofar as the theory provides a physically significant motivation for leaving out part of the standard dynamical picture. In the present case, if RBM would recover the sector of (\ref{BM}) with (\ref{sch}) given by $\hat{H}\Psi=0$, it would be a welcome and relevant result. This is because having accorded to the universal wave function a law-like status, we would expect it not to change over time. However, all of this remains mere speculation as long as a concrete model along these lines is not implemented.

\section{The Metaphysics of Relational Bohmian Mechanics}\label{metrbm}

\subsection{Ontology}
Although the technical implementation of the non-relativistic particle dynamics of RBM is still work in progress, the sketch of such a theory developed in the previous section is physically informed enough to be the object of a fruitful philosophical analysis. In particular, it would be of enormous interests for philosophers to dig into the metaphysics of RBM.\\
With this respect, the first point to highlight is that RBM succeeds in questioning the fundamentality of the dynamical picture of particles changing position in space at different times. RBM's dynamics talks about sequences of instantaneous particle configurations and not about the temporal development of a ``swarm'' of particles deployed over physical space. Let us discuss in detail what the metaphysical significance of this picture might be.\\
The first step to take is to settle for two key elements postulated by the theory, that is, (i) the stuff that is guided, and (ii) what there is. As regards (i), we can say that in RBM the stuff guided still consists of particles. However, while in BM the (time-dependent) wave function ``choreographed'' the motion of the particles through the guiding equation (\ref{gui}), as resulting from the integral curve of the vector field generated in standard configuration space by (\ref{gui}), in the latter case the new dynamics determines a trajectory in the relational configuration space that in no way can be immediately and univocally ``decomposed'' in single particles' trajectories in spacetime. In the case of RBM, then, the only thing we can say is that the ``choreographic'' role of the (time-independent) wave function consists in selecting universal configurations of particles and ordering them along a trajectory in relational configuration space. Formally, this is achieved by virtue of the fact that the amplitude and the phase of the wave function enter the geodesic principle (\ref{jacbar}), thus determining the geometrical features of the shape space $\mathcal{Q}_{0}$. To put it simply, one of the main aims of the RBM program is to supplant ``choreography'' with geometry according to the minimalist requirement, thus making it simpler to argue against the view that the particles are literally ``pushed'' by some kind of field. However, we cannot fully understand this point as long as we do not clarify what is the meaning of (ii), i.e. how RBM answers the question regarding what there is at the fundamental ontological level.\\
Let us start from what there is \emph{not} according to RBM: there is no set of individual loci perduring in an objective universal time flow (Newtonian absolute space and time), nor there is a subsisting set of places-at-a-time called neo-Newtonian spacetime.\footnote{Actually, there is no consensus over whether the spatial and temporal structures entering the dynamics of BM should be best understood as standard absolute space and time or a neo-Newtonian $4$-dimensional structure. For simplicity's sake, we gloss over this further issue.} All there is consists of bits of matter (particles) standing in spatial relations among them: take, say, $N$ particles, arrange them in an array of spatial relations, and you get a point in $\mathcal{Q}_{0}$. This point does not represent a snapshot of a universe with a swarm of $N$ particles in space at a given time, rather, it \emph{defines} what it means to be ``universal'' and what it means to be ``instantaneous''. The dynamics of the theory, then, establishes an ordering of such configurations in the form of a smooth sequence of points (that is, a curve) in $\mathcal{Q}_{0}$ labelled by an arbitrary monotonically increasing parameter,  such that this curve satisfies the principle (\ref{jacbar}). It is very important to note that such a dynamics is non-local in a clear sense: the occurrence of a universal instantaneous configuration depends on the precedent configuration \emph{as a whole}, that is, there is no way to extract exact (viz. non-approximated) dynamical information from parts of a configuration.\\
The ontology of the theory is now quite clear: at the fundamental level there are particles and an irreflexive and symmetric relation $\mathfrak{R}$ that is spatial in nature, which means that a ``coloring'' positive-real-valued function $f$ can be defined in the domain of $\mathfrak{R}$ such that, for each couple of relata, it assigns a value empirically interpretable as a Euclidean distance.\footnote{Of course, such a function is not unique nor objective, since it depends on an arbitrary fixing of the spatial scale.} This relation makes it also possible to define a notion of ``coexistence'': two particles $a$ and $b$ are coexistent just in case $a\mathfrak{R}b$. The notion of configuration can be thus clarified in terms of coexistence, in the sense that a configuration is nothing but a set of coexisting particles. Instead, the dynamical path established in shape space can be interpreted as a strict ordering relation $\mathfrak{C}$ among configurations. This clarifies why, also in the relational case, there is no need in the ontology for the wave function as a concrete physical object. The dynamical picture, in fact, can be either interpreted in Humean terms, hence taking the best-matched stack of configurations as a mosaic on which the dynamical laws supervene, or in a modal realist fashion, claiming for example that each configuration in a curve is a causal structure possessing the power to bring about the subsequent one. Note that, while the modal realist would naturally interpret $\mathfrak{C}$ as some sort of causal linkage among the configurations, the Humean cannot accord such an ontological status to the relation, although, in order to make sense of the mosaic in pre-spatiotemporal terms, she still needs to accord $\mathfrak{C}$ some degree of reality. Furthermore, since in RBM the role of the wave function in generating the dynamics is encoded in the geometrical features of shape space, the absence of commitments to the wave function as a real object can be translated in this context as the absence of commitments to $\mathcal{Q}_{0}$ as a real fundamental space. Under this light, Barbour's quotation at the end of section \ref{besme}, assumes a new and more intriguing meaning. The scale invariance requirement in the Bohmian case (i.e. $\mathfrak{V}\neq0$) can be implemented even in the absence of classical forces ($V=0$) as long as the particles exhibit a quantum behavior ($\mathcal{V}\neq0$). The case where classical forces ``cancel out'' the quantum behavior is instead forbidden.\\
Let us now consider in more detail the appearance of space and time from this picture.
From what has been showed in the previous sections, it is clear that both concepts are reduced in a strong formal sense to the fundamental structures posited by the theory. However, RBM does not fully dispense with spatial and temporal concepts at the fundamental level, although their characterization is sensibly weaker than those we are accustomed to in BM. As regards space, we saw that the fundamental relation taking particles as relata is still spatial in nature, since it can be used to characterize a shape, so we can say that at least a conformal structure is still postulated. In the case of time, instead, we notice that (i) the usual notion of instantaneous configuration can be reduced to that of coexisting particles, and (ii) that $\mathfrak{C}$ provides a strict ordering for configurations. Hence, we are here confronted with an ordering of ``instants'' which is very similar to a B-series of time\footnote{This terminology is of course borrowed from \cite{6}.} for two main reasons. The first is that, due to the monotonicity requirement for $\lambda$, in this picture there is a clear sense in which a given instantaneous configuration comes \emph{after} a precedent one: this ordering plus a choice of a privileged parametrization of the dynamical curve is the supervenience basis on which the appearance of universal time is grounded. The second is that such a picture denies temporal becoming since there is nothing coming to and passing from existence: all there is is a sequence of configurations which can be seen either as a bare ``Humean block'' or as encoding some modal facts of the matter such as ``the subsequent configuration would not have existed having the precedent been different''. It is interesting to note that the metaphysical picture just discussed echoes the Leibnizian view of space as the order of coexisting things, and time as a successive order of things: in a sense, the relations $\mathfrak{R}$ and $\mathfrak{C}$ represent an implementation of these ideas.\\
Together with the afore-mentioned metaphysical basis for the appearance of time in this context comes a notion of particle's identity over time. Actually, the characterization of a configuration as a set of coexisting particles just requires a primitive notion of numerical identity in order to make particles weakly discernible under $\mathfrak{R}$. Such a notion of discernibility is of course needed when two shapes are best-matched, because this procedure is basically an attempt to make two shapes overlap particle by particle. Note, however, that at this stage we are not forced to claim that two juxtaposed particles represent the \emph{same} particle. Once the dynamical best-matching procedure is finally carried out in accordance to the geodesic principle (\ref{jacbar}), and a set of juxtaposed shapes is stacked into a curve in $\mathcal{Q}_{0}$, then we can apply the above discussed metaphysical account for the appearance of a temporal ordering among configurations: it is exactly this derived ordering to ground the notion of spatiotemporal trajectory and, hence, that of particle's identity over time.\footnote{Formally speaking, this amounts to recovering (\ref{qnw}) from the RBM's version of (\ref{eulag}).}\\
To summarize, RBM replaces the usual Bohmian commitment to fundamental entities being particles, absolute space, and absolute time, with a more parsimonious ontology of particles plus two fundamental relations $\mathfrak{R}$ and $\mathfrak{C}$. In the RBM case ``what is guided'' and ``what there is'' can be taken as synonyms, since there are no elements of the primitive ontology that do not enter the dynamical evolution. In this sense, the RBM program can be taken as a recipe to construct a genuinely background independent theory.\\
An open metaphysical problem stemming from the above analysis is what kind of metaphysical priorities we should assign to objects (particles) and relations. With this respect, we suggest that the ontology of RBM is best understood in moderate ontic structuralists terms. Quoting \cite{182}:
\begin{quote}
According to this position, neither objects nor relations (structure) have an ontological priority with respect to the physical world: they are both on the same footing, belonging both to the ontological ground floor. It makes no sense to assign an ontological priority to objects, because instead of having fundamental intrinsic properties, there are only the relations in which they stand. In other words, an object as such is nothing but that what bears the relations. As regards the relations, it makes no sense to attribute an ontological priority to them, for at least insofar as they exist in the physical world, they exist as relations between objects. In sum, as far as the physical world is concerned, there is a mutual ontological as well as conceptual dependence between objects and structure (relations): objects can neither exist nor be conceived without relations in which they stand, and relations can neither exist in the physical world nor be conceived as the structure of the physical world without objects that stand in the relations.\\
(\emph{ibid.}, pp. 31,32)
\end{quote} 
In the case of RBM, according - say - an ontological priority to particles over $\mathfrak{R}$ would not explain why some of them coexist in a configuration and some other of them coexist in another one; that is, there would be nothing inherent into the single particles that would explain this diversification in different configurations. On the other hand, claiming that $\mathfrak{R}$ is prior to particles would raise the question of what would make configurations \emph{physical} as opposed to mere abstract structures; of course, requiring that a physical relation should take concrete objects (in fact, particles) as relata, would answer the question, but this move is obviously precluded to a proponent of this radical form of structuralism. In both cases, then, a moderate form of structuralism would defuse the objections. Moreover, taking configurations as concrete structures would help grounding the non-local dynamical behavior encoded in $\mathfrak{C}$ in a ``holistic'' causal property that is borne by configurations as a whole, being it obviously unexplainable in terms of intrinsic properties of particles (otherwise, there wouldn't be any real non-locality involved).

\subsection{Time from the Quantum}\label{tfq}
The above metaphysical analysis of RBM shares a lot of traits with that of a best-matching theory of classical particles. This is a key point we want to stress: exactly like BM, RBM represents a quantum theory that brings ``classical terms'' in the equations, and such an ontological clarity makes it possible to overcome a very important problem in quantum relational physics. One of the most important results that the RBM program promises to deliver is, in fact, that this theory would yield a well-established and easy to interpret mechanism which provides a fundamental ordering of instantaneous configurations in a quantum context. Indeed, in the ``standard'' relational context, recovering even a weak notion of time (or the appearance thereof) from the Wheeler-DeWitt-like equation (\ref{wdw}) \emph{alone} is quite difficult. Let us try to assess two philosophically interesting proposals for the emergence of time from a quantum relational context,\footnote{Which by no means exhaust the list of strategies for accounting for time in a quantum relational setting. \citet[][sections 20-26]{443} gives a detailed overview of the state-of-the-art in this field: interestingly enough, primitive ontology approaches seem not to be particularly considered in current research, which further motivates the present article.} and point out how RBM provides a better framework for the explanation of the appearance of time from the quantum level.\\
The first proposal is due to Barbour himself and involves the notion of ``time capsule'' \citep{136}.\footnote{It is worth noting that Barbour's proposal was made at a stage in which there was not enough knowledge of shape spaces' geometry (i.e. before work like in \citealp{446}, was carried out). For this reason, we intend Barbour's proposal as a heuristically presented possibility with no detailed mechanism offered (see \citealp{460}, for a discussion on the general approach to records theory in physics). However, what interest us here are the metaphysical implications of Barbour's proposal, independently on its actual (or possible) technical implementation.} For Barbour, the fact that the quantization of a classical relational theory leads to a frozen dynamics in terms of a Wheeler-DeWitt-like equation is a strong hint of the fact that a quantum relational theory must be timeless simpliciter, i.e. it must not allow for whatever ordering, as weak as it might be (let alone, of course, temporal becoming). In order to provide a consistent story of how a quantum relational theory works and how we get the impression of there being change in time, Barbour focuses on the relational configuration space $\mathcal{Q}_{0}$ and accords actual existence to the whole space. What exists is not an actual history (or some collection of consistent histories), but a plurality of ``nows'' as given by instantaneous universal configurations of particles. It happens that some of these nows are structured so that they seem to contain ``records'' of other nows. Just to clarify the ideas, think of a series of footsteps on the sand: this is intuitively a record of someone having walked on the beach. By the same token, there might be a now in which a bunch of particles are mutually arranged in a cloud-chamber-like configuration with some $\alpha$-like tracks in it: this might suggest that this now contains a record of another one which is identical to the former except for the fact that, in the chamber, there is a radioactive atomic nucleus which is likely to perform $\alpha$-decay. However, unlike our intuitive notion of record as some physical consequence of certain past conditions, the records in Barbour's framework are just some sort of suggestive similarities that happen to hold between nows, without any real link connecting these two configurations. Under this view, we basically ``live'' in a time capsule so complex that the particles are arranged in brain-experiencing-temporal-becoming-like configurations. The role of the wave function in this picture is to assign the highest quantum-mechanical probability to those configurations that seem to encode records of the past. Such a mechanism is reminiscent of the analysis made in \cite{404} of the formation of tracks in a cloud chamber due to the $\alpha$-decay of an atomic nucleus. The problem considered in the paper is the following: if the wave function of an $\alpha$-particle being emitted by the nucleus is spherically symmetric, how can it be possible that the interaction with the atoms in the chamber produces a straight track? To cut the story short, Mott described the physical setting in terms of a time-independent Schr\"odinger equation and he was able to show that such a description assigns the highest probabilities to ionization patterns being straight lines. Barbour aims at extending such an analysis to relational configuration space and adding the caveat that the records encoded in the ``privileged'' configurations are not in fact consequences of ``previous happenings''.\\
Barbour's conceptual account of time capsules is complex to spell out in detail (we recommend \citealp[][especially section 3]{402}, and \citealp{403}, for a thorough philosophical discussion of Barbour's views) but the main idea is clear: all possible universal instantaneous configurations are equally real and there is no thing such as a history, i.e. a curve connecting a subset of them. There are many objections - epistemological and metaphysical - that can be raised against Barbour's account of time capsules. Here we would like to consider just one of them, which has an immediate connection with RBM, and takes the form of a very simple question: what is the status of the wave function in the time-capsule picture? If the answer is that the wave function is merely an assignment of probabilities to time capsules, then the subsequent question is: probabilities for what? According to all the major physical accounts of probability, in the world there are not probabilities\footnote{Or degrees of belief, in the Bayesian case.} simpliciter but probabilities for something \emph{to happen}: setting aside probabilities in classical physics, even in the Copenhagen interpretation of quantum mechanics the probabilities associated to the wave function are probabilities of measurement outcomes to occur. So again, what is the meaning of the wave function in Barbour's timeless picture? For sure, it is not the probability of a given time-capsule to be actualized, since \emph{all} the time capsules are equally real. Even if the time capsules bearing the highest probability amplitudes would be more ``special'' in some respect to the others, e.g. by containing both extremely fine structured record-like-configurations and also brain-experiencing-temporal-becoming-like configurations, still it is totally unclear what the higher probability assigned to such complex nows would amount to, given that there is nothing external to these configurations against which we can evaluate their ``likelihood''.\\
All the above problems of course vanish in RBM. First of all, the fact that the theory allows for continuous histories in relational configuration space makes it possible to dispense with the notion of time-capsules: the appearance of records in a given configuration \emph{is} physically linked to the structure of the precedent one in a given history. Moreover, in this framework, the wave function has a clear job, i.e. fixing a history. There are not even such things like superpositions of histories since, once the initial conditions are fixed, a single history is automatically selected. In short, RBM gives a more nice and well-behaved account of the wave function, and provides a quite intuitive mechanism for the appearance of temporal becoming from the underlying quantum regime. Still one may claim that Barbour's picture is more faithful to the minimalist requirement, since it totally eschews \emph{whatever} temporal structure from the ontological picture. This is fair enough, but it seems to us that a metaphysics that renounces time completely would never reach the explanatory power of one that acknowledges the fundamental existence of some time-like ordering, as weak as this might be. Of course, we are are ready to withdraw this claim if confronted with a convincing counter-example.\\
We now turn to a second proposal, due to \cite{405}, for recovering time in a quantum relational context. Roughly, the authors argue that quantizing classical relational systems using Dirac quantization and other derivative approaches is misleading and leads to theories that are not genuine quantizations of the starting ones. In the standard treatment of Hamiltonian gauge systems, in fact, the Hamiltonian constraints are seen as generating merely gauge transformations; once such a framework is quantized, it leads to a dynamics dictated by a Wheeler-DeWitt-like equation, which in turn leads to all the well-known conceptual problems related to the timelessness of such an equation. More precisely, what is lost in the quantization procedure is the possibility to fix a particular parameter in the equations of motion such that a universal time can be shown to emerge. The starting point of the authors is the result due to \cite{406} that, for systems whose dynamics is described by a Jacobi-like principle, the Hamiltonian constraints generate a dynamics that is not just gauge.\footnote{See also \cite{424} for a more general discussion of the problem.} The authors exploit this fact in the context of a path integral approach to the quantization of classical systems\footnote{But see \cite{430} for a new implementation of this framework in terms of a generalized Hamilton-Jacobi formalism.} to show that the global Hamiltonian can be decomposed in a way that gives rise to a universal Schr\"odinger-like evolution for the quantized theory in terms of a privileged parameter that plays the role of an absolute time. At the root of this framework lies an alternative scheme for classifying symmetries, which qualifies classical reparametrization invariance as a kind of symmetry quite distinct from mere gauge.\footnote{This alternative taxonomy is presented and explained in a philosophical fashion in \citet[][section 2]{427}.} This means that the temporal degree of freedom is not an ``otiose'' variable that should be eliminated in the quantization procedure by imposing that the proper operator (the Hamiltonian) annihilates the physical states, but represents an underlying ordering of states that should be preserved in the quantized theory.\\
This second proposal for recovering a notion of time from the relational quantum formalism seems more satisfactory than Barbour's one, since it dispenses with the problematic notion of time capsules. The physical picture provided by Gryb and Th\'ebault is akin to the one presented in this paper because the idea of constructing a universal clock for the subsystems of the universe derives entirely from the fact that the global dynamics unfolds according to an arbitrary monotonically increasing parameter that labels states: hence we have also here an ordering that resembles a B-series of time. However, such an approach reintroduces a universal Schr\"odinger-like dynamics which, from a Bohmian perspective, is less desirable than a Wheeler-DeWitt-like one featuring a static wave function. Moreover, Gryb and Th\'ebault's framework retains a purely quantum spirit in that it just deals with quantum states that are in general superposed, and that are subjected to collapse upon ``extra-universe observation'' (whatever this might mean). In short, this approach exhibits all the conceptual drawbacks of standard quantum theory - starting from a ``cosmological'' measurement problem -, even if it dispenses with the conceptual pain of having a universal dynamics dictated by a Wheeler-DeWitt-like equation alone. In the RBM program, by contrast, there are no troubles related to superpositions and collapses. RBM, in fact, deals with concrete structures, namely, universal configurations of particles, whose dynamics is fixed once and for all when the initial conditions are given. Moreover, RBM is able to recover time from the underlying quantum regime without modifying the assumptions behind the appearance of a Wheeler-DeWitt-like dynamics.

\subsection{Relational Bohmian Mechanics and Local Beables}
Let us finally focus on another delicate aspect of the RBM program, that is, its closeness to the aims of primitive ontology approaches to quantum physics. As we have already pointed out, RBM dispenses with the notion of entities localized in a background spacetime as fundamental. The threat behind such a remark is evident: it seems that RBM betrays the very spirit of primitive ontology approaches to quantum theories discussed in the first section. In fact, postulating material stuff decorating spacetime is the key move in restoring a robust link between quantum phenomena and experiments by unifying the ontological picture: all there is, at whatever scale, is just stuff in spacetime and so things like pointers or spots on a photographic plate are nothing but conglomerates of primitive stuff. Hence, the worry that, by removing space and time from such a picture, we undermine the robustness of this link, becomes all the more justified.\\
First of all, we notice how the derivation of macroscopic objects from spatiotemporally localized primitive stuff - or local beables - is \emph{physically salient} in the sense introduced by \citet[][p. 3161, last paragraph]{228}. Put it simply, the reconstruction of, say, pointer positions from particles dynamics is not only mathematically well-established but also \emph{conservative} as much as physics is concerned: at both levels we have stuff in spacetime; what we are doing is just a coarse-graining of the description. Hence, if we take measurements as physically salient, then there is no problem in arguing that the underlying structure posited by the theory is physically salient as well, and viceversa. But what can we say of a well-defined mathematical procedure to derive stuff in spacetime from fundamental entities that, by themselves, are not in space and time? To shoot it straight on target: even if best-matching lets us derive the empirical predictions of BM from RBM, what is the physical meaning we should attach to this story, provided there is any?\\
The above question is taken up by \citet[][see especially the discussion in section 3]{314},\footnote{These authors consider the issue in the context of quantum theories of gravity that do not posit space and time as fundamental entities; however, their reasoning easily applies to the program considered here.} who point out the two ``directions'' from which Maudlin's worry about physical salience can be considered, namely, ``from below'' (take for granted the physical salience of our physical theory and ask what formal reconstruction of macroscopic objects preserve such trait), and ``from above'' (take for granted that the empirical realm is physically salient and ask how such salience is inherited by our physical theory). Let us assume that RBM has the physical salience we want and ask ourselves how standard BM can inherit such a salience. The answer to this question is indeed very simple: Agreed, RBM does not postulate a background spacetime, but still it postulates a weak spatial ordering for particles and a weak temporal ordering for universal configurations. Moreover, the formal reconstruction of BM from RBM just consists of adding further physical degrees of freedom to such fundamental orderings. Metaphorically speaking, in passing from RBM to BM, we are not altering the structure of the picture: we are just ``embellishing'' it. The same reasoning applies to the inverse problem: if BM is physically salient, then, since the spatial and temporal metrics in BM are supervenient in a very strong sense on (indeed they are formally \emph{reduced} to) weak spatial and temporal orderings in RBM, also this latter theory can count as physically salient. To sum up, there is no real ontological discontinuity between RBM and BM in that (i) both theories rely on spatial and temporal connotations in order to characterize particles' dynamics and (ii) the (neo-)Newtonian background of BM is reduced through a physically justified procedure to the weak orderings of RBM. Moreover, in the case of RBM, the two meanings of primitive ontology considered in the first section simply overlap: all there is consists of particles arranged in configurations through spatial relations and such configurations are exactly what is guided. For this reason, RBM is not only loyal to the primitive ontology spirit but it also offers a more parsimonious and compact account of primitive ontology than BM.

\pdfbookmark[1]{Acknowledgements}{acknowledgements}
\begin{center}
\textbf{Acknowledgements}:
\end{center}
We are very grateful to an anonymous referee, Sean Gryb, and Roderich Tumulka for detailed comments on an earlier draft of this paper. Antonio Vassallo acknowledges support from the Swiss National Science Foundation, grant no. $105212\_149650$.\\

\bibliography{biblio}
\end{document}